\documentclass[12pt]{spieman}  
\usepackage{amsmath,amsfonts,amssymb}
\usepackage{graphicx}
\usepackage{setspace}
\usepackage{tocloft}
\usepackage{textcomp}
\usepackage{tocloft}
\usepackage[flushleft]{threeparttable}
\usepackage{graphicx}
\usepackage{lipsum}
\usepackage{float}
\usepackage{amsmath}
\usepackage{latexsym}
\usepackage{booktabs,etoolbox}
\usepackage[utf8x]{inputenc}
\usepackage{caption}

\title{Design and modeling of a tunable spatial heterodyne spectrometer for emission line studies}

\author[a,b]{Nirmal Kaipachery}
\author[a]{Sridharan Rengaswamy}
\author[a]{Sripadmanaban Sriram}
\author[a]{Jayant Murthy}
\author[a]{Suresh Ambily}
\author[d]{Margarita Safonova}
\author[a,c]{Aickara Gopinathan Sreejith}
\author[a]{Joice Mathew}
\author[a]{Mayuresh Sarpotdar}

\affil[a]{Indian Institute of Astrophysics, Bangalore, India, 560034}
\affil[b]{University of Calcutta, Kolkata, India, 700073}
\affil[c]{Space Research Institute, Austrian Academy of Sciences, Schmiedlstrasse 6, 8042 Graz, Austria}
\affil[d]{M. P. Birla Institute of Fundamental Research, Bangalore, India, 560001}

\cftpagenumbersoff{figure}
\cftpagenumbersoff{table} 
\begin{document}
\maketitle

\begin{abstract}
Spatial Heterodyne Spectroscopy (SHS) is a relatively novel interferometric technique similar to the Fourier transform spectroscopy  with heritage from the Michelson Interferometer. An imaging detector is used at the output of a SHS to record the spatially-heterodyned interference pattern. The spectrum of the source is obtained by Fourier transforming the recorded interferogram. The merits of the SHS -- its design, including the absence of moving parts, compactness, high throughput, high SNR and instantaneous spectral measurements -- make it suitable for space as well as for ground observatories. The small bandwidth limitation of the SHS can be overcome by building it in tunable configuration (Tunable Spatial Heterodyne Spectrometer, TSHS). In this paper, we describe the design,  development and simulation of a TSHS in refractive configuration suitable for optical wavelength regime. Here we use a beam splitter to split the incoming light compared with all--reflective SHS where a reflective grating does the beam splitting. Hence the alignment of this instrument is simple compared with all--reflective SHS where a fold mirror and a roof mirror are used to combine the beam. This instrument is intended to study faint diffuse extended celestial objects with a resolving power above 20000, and can cover a wavelength range from 350 nm to 700 nm by tuning. It is compact and rugged compared with other instruments having similar configurations.
\end{abstract}


 \\
{\noindent \footnotesize\textbf{*}Corresponding author,  \linkable{nirmalk@iiap.res.in} }

\begin{spacing}{2}   

\section{Introduction}

High-resolution spectroscopy of faint extended objects in UV and visible is difficult because one has to compromise between the sensitivity and the resolving power. There are certain objects and phenomena in the Solar System:  comets, nebulae, planetary satellites, planetary auroras and interplanetary medium, that are faint, extended and reservoirs of information. Low-resolution spectroscopy can only reveal their basic parameters such as composition, intensity, and energy distribution. However, high-resolution spectroscopy can reveal additional information such as velocity, temperature, pressure, and isotopic signatures\cite{lyman}. In order to obtain high-resolution spectra of an extended target, interferometric spectral measurement technique such as Fourier Transform Spectrometer (FTS) or Spatial Heterodyne Spectrometer (SHS)\cite{Har} can be used. 

Since these instruments do not have slits the throughput of these instruments is higher than that of slit based grating spectrometer. However, off axis rays in the input beam, reduces the fringe visibility and the efficiency of the system which in turn put a 
limitation on the field of view (FOV) solid angle of the SHS. The FOV of an uncompensated\cite{harland92} (without a field-widening prism) SHS is relatively small, given by $\frac{2\pi}{R}$, where $R$ is the resolving power of the instrument. This FOV can be increased by introducing the field-widening prism in SHS arms and, hence, they are more sensitive compared to the grating spectrometers with similar configurations.

FTS instruments are easier to work with in the long wavelength regime (infrared) of the electromagnetic spectra because the maximum sampling interval in OPD (optical path difference) space must be $\lambda_{min}/2$\cite{Hariharan}, which is very challenging to achieve at shorter wavelengths (visible and UV). Furthermore, FTSs tend to be bulky. SHS instruments, on the other hand, can be used at short wavelengths, are lightweight, compact, and have flexible alignment considerations\cite{Roes}. Due to these advantages, SHS systems can be used to study H$\alpha$ emission from star-forming regions, atmospheric emission lines, and emission lines from planetary auroras. 

The tunable SHS system described in the literature\cite{dawson} are in all--reflective configurations, which requires a roof mirror and a fold mirror, and hence are suitable for UV missions. In this paper, we describe modeling, simulation, and the design of a tunable SHS instrument in refractive configuration for the optical wavelength regime, where we use a beam splitter for splitting the incoming beam. Hence, the alignment of this instrument is simpler, compared with all-reflective SHS where fold and roof mirrors are used. This instrument can be used to observe and study faint extended emission line targets by retrieving the high-resolution spectra (with resolving power greater than 20000) from the entire source. In the near future, we will use this instrument to observe the H$_{\alpha}$ emission lines from Be stars to study their shape and variability. Another science case of this instrument is to monitor the flaring activities of red dwarf stars in H$\alpha$ by looking at the star continuously through a meter-class telescope.

\section{Fundamentals of SHS theory}

\begin{figure}[h!]
\centering
\includegraphics[scale=0.6]{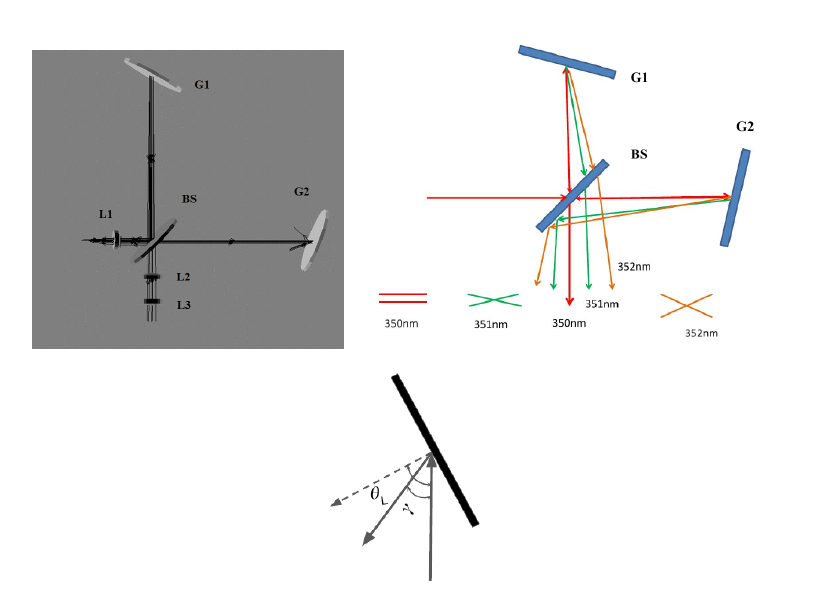}
\caption{{\it Top Left:} The schematic diagram of the SHS system. L1, L2, and L3 are lenses, G1 and G2 are the gratings, and BS is the beam splitter. {\it Top Right:} The concept of fringe formation in SHS. The Littrow wavelength 350 nm doesn't create any fringes because the waves emerge parallel to each other from the two arms of the interferometer. Wavelengths other than Littrow form fringes on the detector because they are tilted by the gratings. {\it Bottom:} Zoomed version of one of the gratings.} 
\label{fig:shs_wave}
\end{figure}

SHS instruments, like the FTS, share structural similarities with the Michelson interferometer. The primary difference between the two is that in the SHS the mirrors  are replaced with the gratings (G1 and G2 in Fig.~\ref{fig:shs_wave}). The position of the gratings is fixed such that the path difference between the two arms is zero, with the tilt of the gratings set for a particular central wavelength called the Littrow wavelength ($\lambda_L$). The heterodyne wavenumber is defined as $k_L = 1/\lambda_L$. Wavelengths near the Littrow wavelength form fringes in the detector plane because the grating introduces a tilt on the wavefront proportional to the absolute deviation between the wavelength and the Littrow wavelength, as shown in Fig.~\ref{fig:shs_wave}.

In order to understand the system mathematically, we have considered one grating in the SHS configuration (Fig.~\ref{fig:shs_wave}, {\it Bottom}). The relation between the incident ray and the refracted ray can be written using the grating equation,
\begin{equation}
d\left(\sin{\theta_{in}} + \sin{\theta_{diff}}\right) = m\lambda\,,
\label{eq:grat}
\end{equation}
where $d$ is the pitch of the grating, $\theta_{in}$ is the incident angle, $\theta_{diff}$ is the diffracted angle, $m$ is the order of grating, and $\lambda$ is the wavelength of the light. Hence, for the grating shown in Fig.~\ref{fig:shs_wave}, {\it Bottom}, the equation becomes
\begin{equation}
d\left(\sin{\theta_{L}}+ \sin(\theta_{L}-\gamma)\right) = m\lambda\,,
\label{eq:theta_l}
\end{equation}
where $\theta_{L}$ is the Littrow angle for the wavelength $\lambda_L$, and $\gamma$ is the angle between the incident and diffracted light. The incident beam and the diffracted beam follow the same path in the Littrow configuration, and $\gamma= 0$. Eq.~\ref{eq:theta_l} then becomes Eq.~\ref{eq:sin_thetal}, which gives the relation between Littrow wavelength and Littrow angle,
\begin{equation}
\sin(\theta_{L}) = {\frac{m}{2k_Ld}}
\label{eq:sin_thetal}\,,
\end{equation}
where $k_L = 1/\lambda_L$ is the wavenumber in cm$^{-1}$. Expanding the sine terms in Eq.~\ref{eq:theta_l} using the small angle approximation for $\gamma (\sin{\gamma} = \gamma, \cos{\gamma} = 1)$, and eliminating $d$ and $m$ using Eq.~\ref{eq:sin_thetal}, we obtain
\begin{equation}
\gamma = \frac{2(k-k_L)\tan(\theta_L)}{k}
\label{eq:gamma}\,,
\end{equation}
where $k = 1/\lambda$ is the wavenumber in cm$^{-1}$. If two beams with the wavelength 1/{\it k}, tilted at angles $\gamma$ and $-\gamma$, respectively, are made to interfere each other, they will produce a fringe pattern with a spatial frequency of $2k\sin{\gamma}$. Thus, the frequency of the fringe pattern is (in small angle approximation of $\gamma$)
\begin{equation} 
f_x = 2k\sin{\gamma} = 4(k-k_L)\tan(\theta_L)
\label{eq:ff}\,,
\end{equation}
where $f_x$ is the frequency of the fringes known as the Fizeau fringe frequency.
In interferometers, the intensity pattern and the fringe frequency are related through the following equation, where $x$ is the pixel position along the $X$ axis of detector, considering detector is in $XY$ plane,
\begin{equation}
I(x) = I\left(1+\cos(2 \pi f_x x)\right)\,.
\label{eq:inter_pattern}
\end{equation}

Equation.~\ref{eq:inter_pattern} shows no differentiation between wavelengths that are lower or higher than heterodyne wavelength since interference pattern depends upon $|k-k_L|$. This can be mitigated by slightly tilting the grating by an angle $\frac{\phi}{2}$ in a direction perpendicular to the plane of interference, which creates a vertical fringe pattern with frequency\cite{dawson} $f_y = \phi k$ and has a dependency on wavelength. This is explained in detail in Section~\ref{sec:tilt}. 
Therefore, the fringe pattern can be represented as a function of $x$ and $y$ for SHS system as in Eq.~\ref{eq:fringe}. For the more general case, where the incoming wave consists of more than one wavenumber, we use Eq.~\ref{eq:intensity},
\begin{eqnarray}\label{eq:fringe}
&&I(x) = B\left( 1+\cos{[8 \pi (k-k_L) \tan{\theta_L}x + 2\pi\phi ky]}\right)\,,\\
&&I(x) = \int_{-\infty}^{\infty} B(k) \left(1+\cos{[8 \pi (k-k_L)}\tan{\theta_L} x + 2\pi\phi ky]\right)dk\,.
\label{eq:intensity}
\end{eqnarray}
Eq.~\ref{eq:intensity} indicates that the intensity distribution on the detector is the Fourier cosine transform of the spectra. Hence, the spectra can be retrieved by taking the inverse Fourier cosine transform of the generated interferogram. Since the system uses the entire grating, the resolving power $R$ of the instrument is 
\begin{equation}
R = \frac{2 \times m \times W}{d},
\label{eq:r}
\end{equation}
where $W$ is the illuminated grating width and $m$ is the grating order.

\subsection{Limitations of SHS}

The SHS has advantages over a grating and Fourier transform spectrometer such as compactness, sensitivity and ruggedness. No slits are used in the SHS allowing the instrument to collect more light, therefore the sensitivity of the instrument is high compared with grating spectrometers. Thus, this instrument is suited well for studies of faint extended sources.  Since the resolving power of SHS depends on the illuminated grating width, grating groove density and order of the grating (Eq.~\ref{eq:r}), larger optics (to accommodate large beam width and higher orders) or gratings with higher groove density are needed for higher resolving power.
Such high resolving power narrows the observable bandpass, which another disadvantage. This can be explained with an example: if the resolving power of the SHS instrument is 24000 at the heterodyne wavelength of 650 nm, then the resolution ($\Delta\lambda$) of the instrument is 0.027 nm (650/24000). Therefore, an input beam having an absolute wavelength deviation $p\Delta\lambda$ from the Littrow wavelength (here 650 nm) should produce $p$ fringes, where $p$ is an integer. If a $1024 \times 1024$ detector is used to image the fringes, then from the Nyquist criteria the maximum number of fringes that can be imaged is 512. The bandpass of SHS can be found as
\begin{equation}
BW = 2 \times \Delta \lambda \times N\,,
\label{eq:shs_bw}
\end{equation}
where $BW$ is the bandwidth in nm, and $N$ is the number of fringes. Hence, the bandwidth of above configuration is 27 nm. This limitation of the SHS can be mitigated by either using a tunable SHS\cite{sona} (TSHS), or a Multi-Order SHS\cite{harl}. 

\section{Modeling of TSHS system in Zemax\textsuperscript{\textregistered}}

\begin{figure}[h!]
\centering
\includegraphics[scale=0.8]{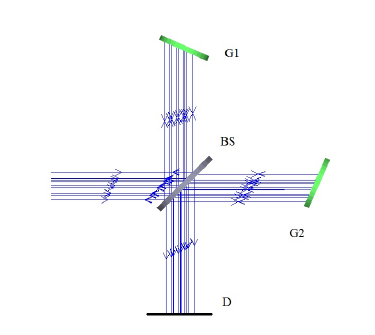}
\caption{A conceptual diagram of a tunable SHS modeled in Zemax\textsuperscript{\textregistered}. G1, G2 are the gratings, BS is the beam splitter, and D is the detector.}
\label{fig:shs_model}
\end{figure}

\begin{figure}[h!]
\centering
\includegraphics[scale=0.4]{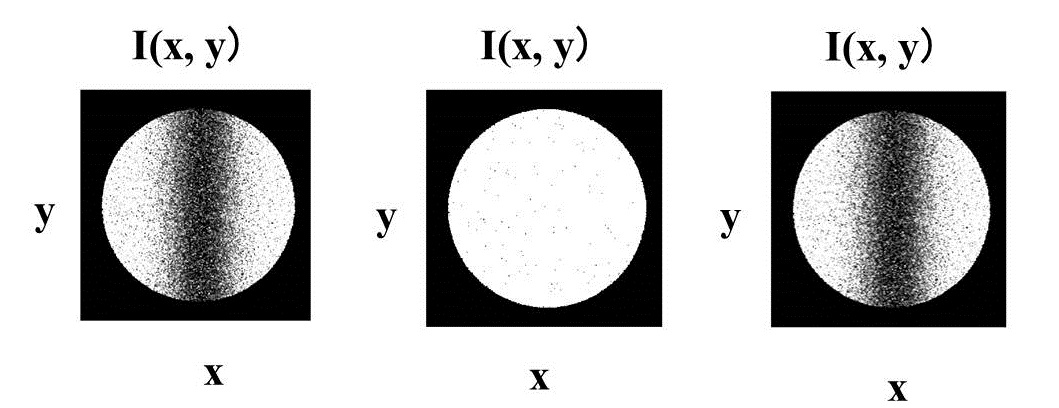}
\caption{The simulated detector view: generated interferogram in $x$ and $y$ positions, $I(x,y)$ is displayed when the incoming beam constituted of only the wavelength 450.0187 nm ({\it Left}), 450 nm (Littrow) ({\it Middle}) and 449.981 nm ({\it Right}). The size of the detector is 12 mm $\times$ 12 mm, having 512$\times$512 pixels. The white circular patch (10 mm beamwidth) is the illuminated area of the detector. These images have been processed to enhance the fringe contrast.} 
\label{fig:initial}
\end{figure}

We have modeled a tunable SHS (Fig.~\ref{fig:shs_model}) using the commercially available optical simulation software Zemax\textsuperscript{\textregistered}. The model includes a beam splitter (BS) where half the light is transmitted and half reflected, two holographic gratings (G1, G2) of groove density (1/pitch) 1200 lines/mm, and a detector (D) of size 12 mm $\times$ 12 mm with 512$\times$512 pixels. Initially the gratings are tilted to a Littrow angle of $15.6643^{\circ}$, which corresponds to a Littrow wavelength of 450 nm. The collimated beam of 10 mm diameter is allowed to fall on the beam splitter. The reflected and the transmitted beams are incident on  the gratings G1 and G2, respectively, and are combined after leaving the beam splitter. Since the incoming beam has only one wavelength of 450 nm, there are no fringes on the detector (Fig.~\ref{fig:initial}, middle panel) as there is no path difference between the rays coming from two arms of the TSHS instrument.

In this configuration, the resolving power $R$ of the system (Eq.~\ref{eq:r}) is 24000, which means that one fringe is generated on the detector for a change of 0.0187 nm $\left(\frac{450}{24000}\right)$ in the wavelength. In order to verify this, we changed the incoming wavelength to 450.0187 nm in Zemax\textsuperscript{\textregistered} model and obtained one fringe (Fig.~\ref{fig:initial}, {\it Left}) on the detector. Similarly, the other fringe is generated on the detector at 449.981 nm (Fig.~\ref{fig:initial}, {\it Right}).

\subsection{Tunable SHS}

\begin{figure}[h!]
\begin{center}
\includegraphics[scale=0.35]{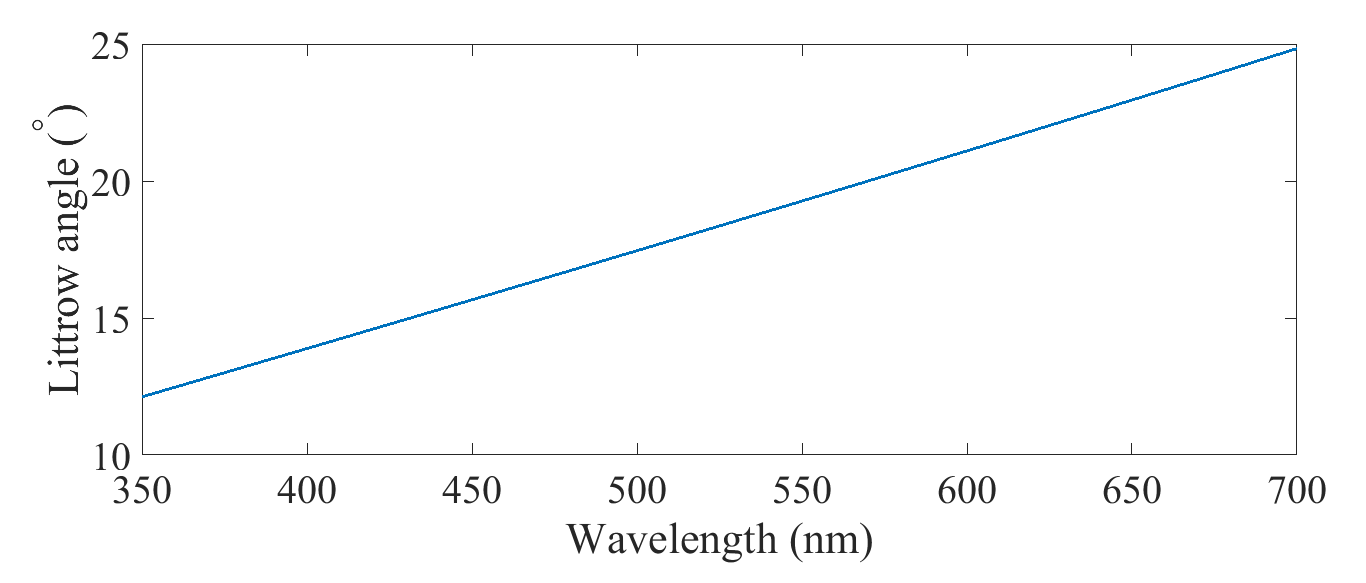} 
\end{center}
\caption{The relation between the Littrow angle and wavelength for the holographic grating with 1200 lines/mm.}
\label{fig:lit_ang}
\end{figure}

Tuning for different wavelengths in the TSHS is carried out by tilting the grating for different Littrow configurations; i.e. changing the Littrow wavelength (central wavelength of the bandwidth). The relation between the Littrow angle and the Littrow wavelength is given by Eq.~\ref{eq:sin_thetal}. We have found the Littrow angle for different heterodyne wavelengths for the grating we have used (holographic grating 1200 lines/mm), and plotted in Fig.~\ref{fig:lit_ang}. In the Zemax\textsuperscript{\textregistered} model, we tilted the grating for different Littrow configuration and obtained the fringes.

\subsection{Tunability of the  system}

\begin{figure}[h!]
\centering
\includegraphics[scale=0.4]{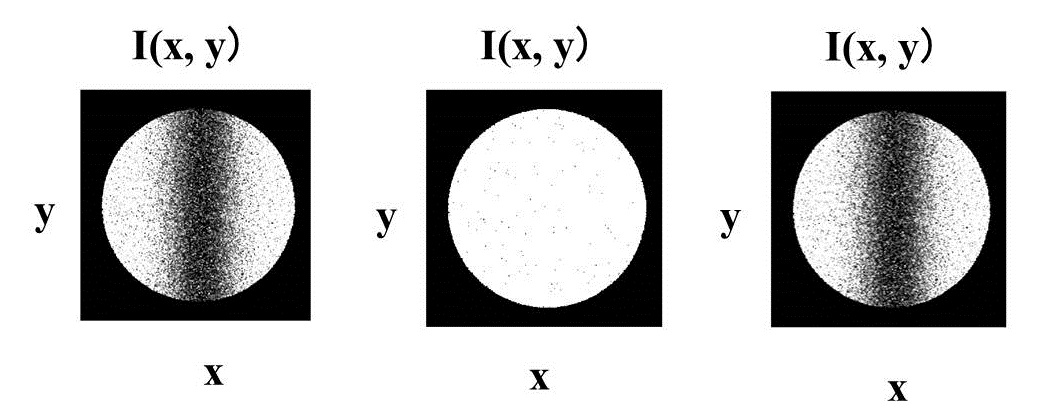}
\caption{Simulated detector view: generated interferogram in $x$ and $y$ positions, $I(x,y)$ is displayed when the input beam constitutes only the wavelength 549.977 nm ({\it Left}), 550 nm (Littrow) ({\it Middle}) and 550.023 nm ({\it Right}). The size of the detector is 12 mm $\times$ 12 mm, having $512 \times 512$ pixels. The white circular patch (10-mm beamwidth) is the illuminated area of the detector. These images have been processed to enhance the fringe contrast.} 
\label{fig:tune}
\end{figure}

In order to check the tunability of the TSHS system the Littrow angles of gratings G1 and G2 were changed to $19.269^{\circ}$, which corresponds to a Littrow wavelength of 550 nm. Since all other configurations were the same, one fringe should be obtained on the detector at a wavelength  change of 0.023 nm. Fig.~\ref{fig:tune} ({\it Middle}) shows the detector view for a Littrow wavelength of 550 nm, with the images for wavelengths of 549.977 nm and 550.023 nm shown on left and right, respectively.

\subsection{Identifying the wavelength}
\label{sec:tilt}

\begin{figure}[h!]
\centering
\includegraphics[scale=0.4]{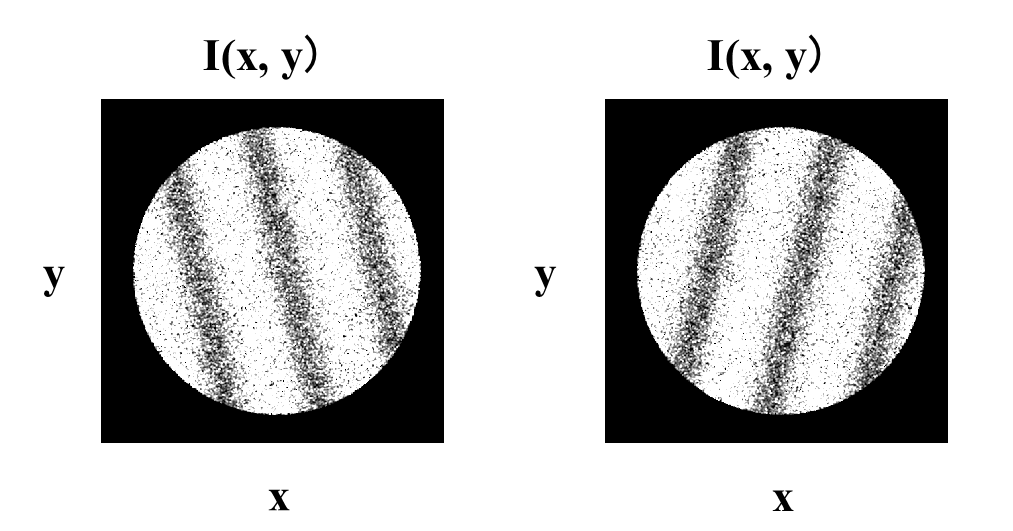}
\caption{Simulated fringes obtained for wavelength 450.0562 nm ({\it Left}) and for 449.944 nm ({\it Right}), when the grating G1 is tilted $0.001^{\circ}$ in the direction perpendicular to the Littrow angle. The Littrow wavelength for this setup is 450 nm. These images have been processed to enhance the fringe contrast.}
\label{fig:id_wave}
\end{figure}

Wavelengths on either side of the Littrow wavelength will generate interferograms with the same spatial frequency: the same number of fringes will be formed by the wavelengths $\lambda_o + \Delta \lambda$ and $\lambda_o - \Delta \lambda$ (Fig.~\ref{fig:initial}). Hence, it is essential to differentiate between fringes formed by $\lambda_o + \Delta \lambda$ and $\lambda_o - \Delta \lambda$. This can be done by tilting the grating in a direction perpendicular to the direction of Littrow angle. This arrangement creates fringes aligned in different direction as in Fig.~\ref{fig:id_wave} for $\lambda_o + \Delta \lambda$ and $\lambda_o - \Delta \lambda$. 

\section{Design considerations of a tunable SHS}

Our TSHS system is designed to work as the back-end instrument for a ground-based telescope in the spectral regime between 350 nm to 700 nm.  We have commissioned this instrument at the 2.34-m Vainu Bappu Telescope (VBT) of the  Vainu Bhappu Ovservatory (VBO)\renewcommand{\thefootnote}{\alph{footnote}}\footnote{https://www.iiap.res.in/centers/vbo}, Kavalur, India. The F-ratio of the beam at the prime focus of the telescope is 3.25, and the light is collected from the prime focus and fed to our system through a 100 $\mu$m diameter optical fiber. We selected two holographic gratings of pitch ($d$) as $1/1200$ lines per mm for the system. The maximum theoretical resolving power of the system was fixed to be 24000, which requires the width of the grating to be illuminated as 10 mm for order $m=1$ (Eq.~\ref{eq:r}). 

\subsection{Entrance Optics and Collimation}

As the minimum expected size of the beam on the grating was 10 mm, we fixed 10 mm as our beam width at the entrance optics, thereby accounting for the beam divergence. The $F$-ratio of the incoming beam is 3, and the focal length $f$ of the collimating lens is found as follows,
\begin{eqnarray}
&&f = F\# \times \Phi \\ \nonumber
&&f = 3 \times 10 = 30 \,\text{mm}\,,\label{eq:11}
\end{eqnarray}
where $F\#$ is the $F$-ratio of the incoming beam, and $\Phi$ is the diameter of the beam.

\subsection{Gratings and beam splitter}

We used a one inch grating operating in the first order ($\pm 1$) and a cube beam splitter of $40\times 40\times 40$ mm size.

\subsection{Fringe localization plane}

The fringe localization plane (FLP) is where the interference fringes can be viewed with maximum clarity. Hence, at this plane the fringes can be obtained at maximum contrast. The distance $z_0$ of the FLP from the beam splitter is found using Eq.~\ref{eq:flp}\cite{Hosse},
\begin{equation}
z_0 = \dfrac{L}{2\cos^2(\theta_L)}\,,
\label{eq:flp}
\end{equation}
where $L$ is the average optical path inside the TSHS. The position of the FLP depends on the source as well as on the instrument. The FLP of a point source is unlocalized and has a greater contrast, compared with the FLP of an extended source which is localized with less visibility. 

\subsection{Exit Optics}

\begin{figure}[htb!]
\centering
\includegraphics[scale=0.45]{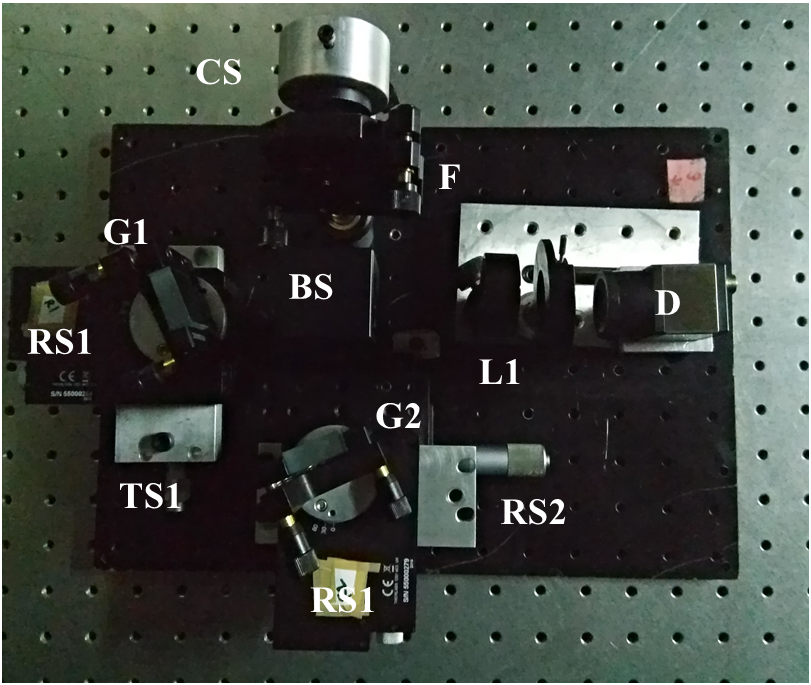} 
\caption{The TSHS setup arranged on a bread board (top view). CS is the collimator setup, F is the filter, L1 is the imaging lens, BS is the beam splitter, D is the detector, RS1 and RS2 are the rotational stages, and TS1, TS2 are the translational stages.}
\label{fig:shs_breadboard}
\end{figure}

The exit optics re-images the FLP onto the detector. We have kept the exit optics on a linear stage whose position can be changed to obtain clear fringes on the detector. The diameter of the beam at the FLP is same as the illuminated grating width. Therefore, a standard 25-mm aperture lens with 50 mm focal length is selected to re-image FLP onto the detector. Our selected components are listed in Table~\ref{tab:components}, and we have completed a lab setup of the system on an optical breadboard (Fig.~\ref{fig:shs_breadboard}).

\begin{table}[h!]
\caption{The components procured for building TSHS instrument.} 
\label{tab:components}
\centering
\begin{tabular}{|l|p{70mm}|}
\hline
\rule[-1ex]{0pt}{3.5ex} Components & Description \\
\hline
\rule[-1ex]{0pt}{3.5ex} Grating &  Holographic gratings, 1200 lines/mm \newline visible (bare aluminum coating)
 \\
\rule[-1ex]{0pt}{3.5ex} Beam splitter & Standard cube beam splitter (50:50) \newline $40\times 40\times 40$ mm  \\
\rule[-1ex]{0pt}{3.5ex} Rotational stages& Newport URS50BPP \newline $0.00022^{\circ} $ resolution \newline $0.0005^{\circ}$ minimum incremental motion.\\
\rule[-1ex]{0pt}{3.5ex} Detector &  Sony IMX249
CMOS camera \newline $1920 \times 1200$
 \newline 8-bit camera \newline  $5.86 \times 5.86\,\mu$m pixels\\
Entrance optics & convex lens $f= 30$ mm \newline 10 mm diameter \\
Exit optics & convex lens $f= 50$ mm \newline 25 mm diameter\\
\hline
\end{tabular}
\end{table}

\section{Experimental setup and first results}

\begin{figure}[htb!]
\centering
\includegraphics[scale=0.8]{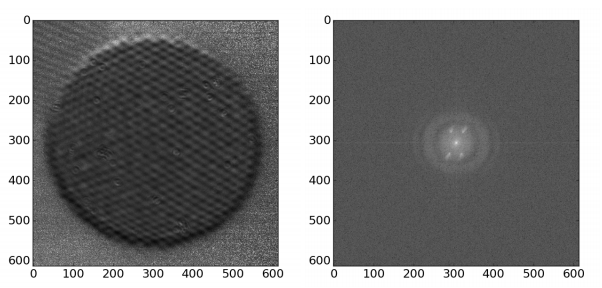}
\caption{{\it Left:} The fringe obtained from the sodium vapor lamp. Two distinct fringes can be seen in the image corresponding to wavelengths 589 nm and 589.6 nm, respectively. {\it Right:} The two-dimensional FFT of the flat-corrected interferogram.}
\label{fig:na_fringe}
\end{figure}

We have used a sodium vapor lamp as our source with the sodium D lines at 589 nm and 589.6 nm, and tuned the system to those wavelengths. The light from the source is fed to our TSHS system through an optical fiber cable of core diameter 100 $\mu$m. A linear stage in each arm of the TSHS to adjust the grating position as well as the OPD. Since the source contains two emission lines, the interferogram also contained fringes of two different spatial frequencies as can be seen in Fig.~\ref{fig:na_fringe}. 

\begin{figure}[htb!]
\centering
\includegraphics[scale=0.8]{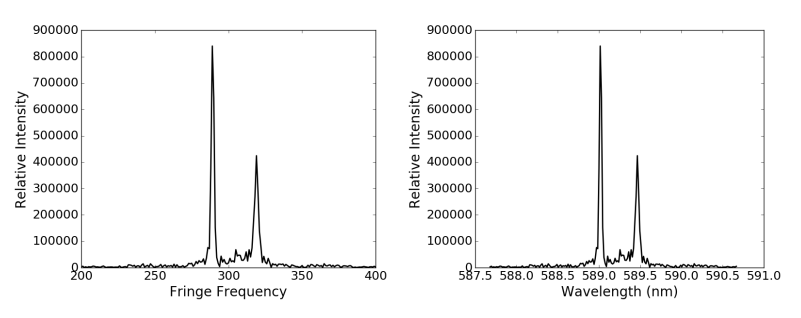} 
\caption{The spectrum of the sodium vapor lamp retrieved from the interferogram, ({\it Top:}) in pixel scale and in ({\it Bottom:}) in wavelength scale.}
\label{fig:na_spectrum}
\end{figure}

The spectrum is retrieved by applying a two-dimensional Fourier transform (Fig.~\ref{fig:na_fringe}) on the interferogram. The spectrum shown in Fig.~\ref{fig:na_spectrum} is obtained by taking a line cut of the two-dimensional interferogram.

\subsection{Estimation of Resolving Power ($R$)}

\begin{figure}[htb!]
\centering
\includegraphics[scale=0.5]{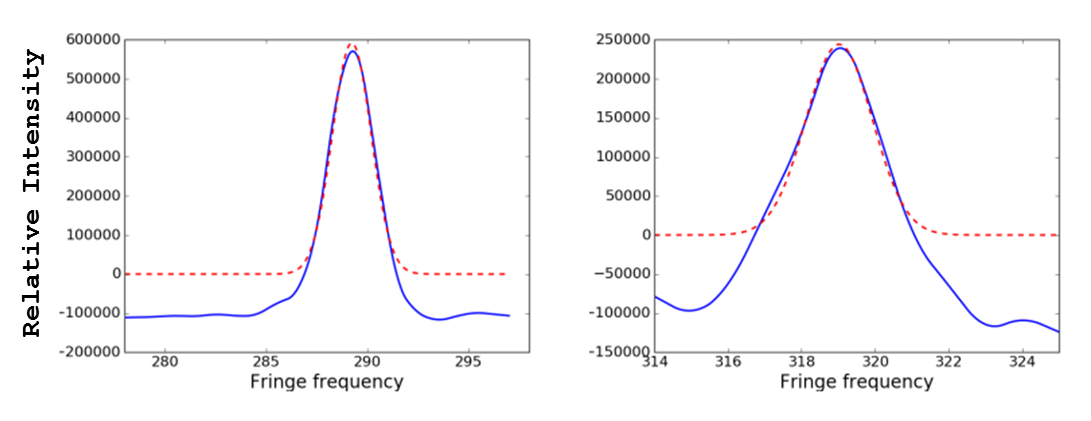}
\caption{{\it Left:} The emission line corresponding to 589 nm (blue color) and the fitted Gaussian curve (red dotted line). {\it Right:} The emission line corresponding to 589.6 nm  (blue color) and the fitted Gaussian curve  (red dotted line).}
\label{fig:spec_fit}
\end{figure}

The two sodium D lines were fitted using Gaussian profiles (Fig.~\ref{fig:spec_fit}, Table~\ref{tbl:fit_para}).  Before the fitting, we have interpolated the data to 1001 points and smoothened using a Savitzky-Golay filter.
The difference in the wavelength of sodium emission lines (here 0.6 nm) corresponds to $289.232-319.019= 29.787$ fringe frequency (resolution elements) (Table~\ref{tbl:fit_para}), and one fringe frequency corresponds to a wavelength of $\dfrac{0.6}{29.787}=0.020$ nm. Therefore, the resolving power of the instrument is $R=\dfrac{589}{0.020}=29450$,  giving the FOV solid angle at the grating ($\Omega$) of the instrument as $2.13\times10^{-4} sr$.

\begin{table}[htb!]
\caption{Parameters obtained from the fitting of emission lines.} 
\label{tbl:fit_para}
\centering
\begin{tabular}{|l|c|c|}
\hline
\rule[-1ex]{0pt}{3.5ex} Parameter & Peak 589 nm & Peak 589.6 nm\\
\hline
\rule[-1ex]{0pt}{3.5ex} Amplitude  &  $1.4226\times 10^6 \pm 2.75 \times 10^4$ & $5.5331\times 10^5 \pm 1.61\times 10^4$ \\
\hline
\rule[-1ex]{0pt}{3.5ex} Central fringe frequency  &   $289.232291 \pm  0.021386$ &
$319.019890 \pm 0.030433$ \\
\hline
\rule[-1ex]{0pt}{3.5ex} FWHM  &  $0.95939241 \pm  0.021386 $ &
$ 0.90398182 \pm  0.030433$ \\
\hline
\end{tabular}
\end{table}

\subsection{Interferogram from a continuum source}

\begin{figure}[htb!]
\centering
\includegraphics[scale=0.35]{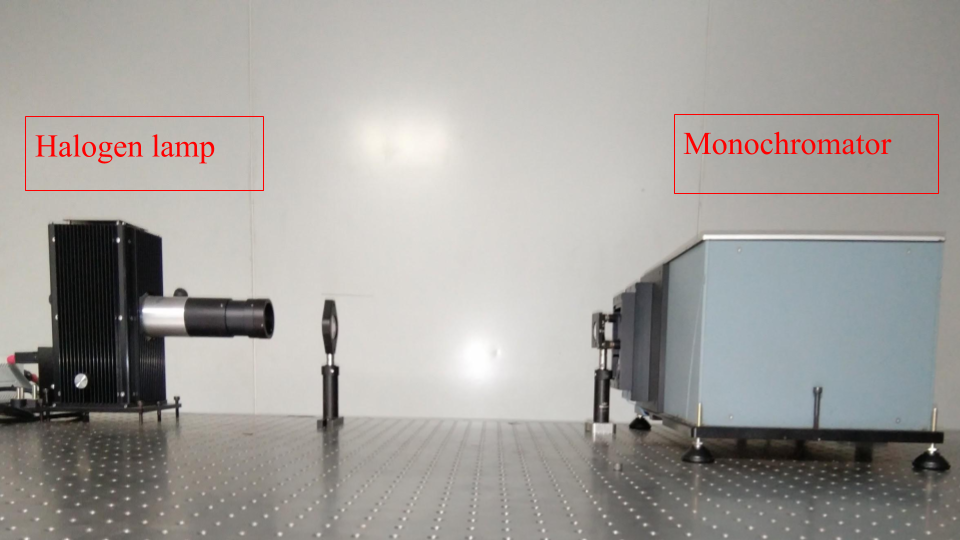} 
\caption{The setup of monochromator Andor SR-303IA and halogen lamp. A halogen lamp produces a continuous spectrum. The monochromator is used to select a particular wavelength region.}
\label{fig:sources}
\end{figure}

\begin{figure}[htb!]
\centering
\includegraphics[scale=0.7]{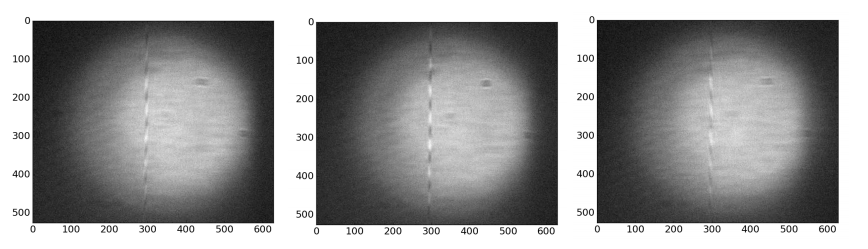} 
\caption{Fringes obtained for a continuum source -- halogen lamp. The central wavelength is changed using the monochromator. {\it Left:} Fringes obtained for a wavelength band centered at 587 nm. {\it Middle:} Fringes obtained for a wavelength band centered at 588 nm. {\it Right:} Fringes obtained for a wavelength band centered at 589 nm. As the wavelength moves away from the central wavelength, there is a change in fringe frequency.}
\label{fig:halogen_fringe}
\end{figure}

 A passband filter is usually used in SHS system to limit the spectrum to one SHS sideband or to limit the spectral range, in order to minimize the shot noise contribution. However, we have not used a spectral filter in our lab setup. Instead, we tested the tunability of the system using a halogen lamp monochromator Andor SHAMROCK–SR-303I (Fig.~\ref{fig:sources}), for which we selected a band around 588 nm. The light from the source is fed to the monochromator and the required wavelength, as well as the bandwidth, are selected by adjusting the monochromator's parameters. It has a F/4 aperture and a focal length of 303 mm, giving a wavelength range from 170 nm to 10 $\mu$m and a wavelength resolution of 0.05 nm. The obtained fringes for this particular setup are shown in Fig.~\ref{fig:halogen_fringe}.  Since the halogen lamp spectrum is a broad-band source, the interferogram is only modulated in a small region around the zero OPD location (Fig.~\ref{fig:halogen_fringe}). The fringes obtained for the wavelength centered at 587 nm are tilted as shown in Fig.~\ref{fig:halogen_fringe} ({\it Left}), while the fringes from the wavelength centered at 589 nm are tilted in the opposite direction (Fig.~\ref{fig:halogen_fringe}, {\it Right}), in agreement with our simulation (Sec.~\ref{sec:tilt}). Fringes obtained for the Littrow wavelength (588 nm) do not show any tilt as expected (Fig.~\ref{fig:halogen_fringe}, {\it Middle}). 

\section{Flat-fielding in SHS}

\begin{figure}[htb!]
\centering
\includegraphics[scale=0.75]{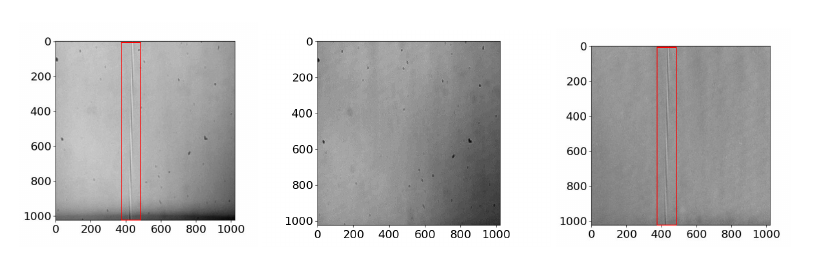}
\caption{{\it Left:} Uncorrected interferogram. {\it Middle:} The master flat. {\it Right:} Flat-field corrected fringes.}
\label{fig:flats}
\end{figure}

The flat-fielding of the instrument is affected by the presence of dust on the optical components as well as by the intrinsic non-uniformity of the detector, and this will be seen in the interferogram. In order to get the flat field (uniform illumination on the detector) after obtaining the fringe, one arm of the interferometer was closed, and this uniform illumination was imaged with the same detector. This process was repeated by closing the other arm of the SHS. A master flat was generated from these flats, and the flat correction was done as explained in  Englert et al. (2006)\cite{flat}. In Fig.~\ref{fig:flats}, we show the original interferogram ({\it Left}), the master flat ({\it Middle}), and the flat-field corrected interferogram ({\it Right}), respectively.

\section{First-light results}

We have observed with the TSHS system using the 2.34-m telescope VBT. The f/3 beam from the prime focus of the telescope is collected through a 100 $\mu$m-diameter optical fiber of approximately 35 m length, and is fed to the TSHS. Prior to the observation, we have aligned the TSHS for the for H$_{\alpha}$ line (656.3 nm) line, since many stars have strong emission line in this wavelength. A 4 \AA\, (full-width-half-maximum) pre-filter is placed between the entrance optics and the beam splitter for enhancing the visibility of the fringes.

\subsection{Effect of vibration on the system}

\begin{figure}[htb!]
\centering
\includegraphics[scale=0.6]{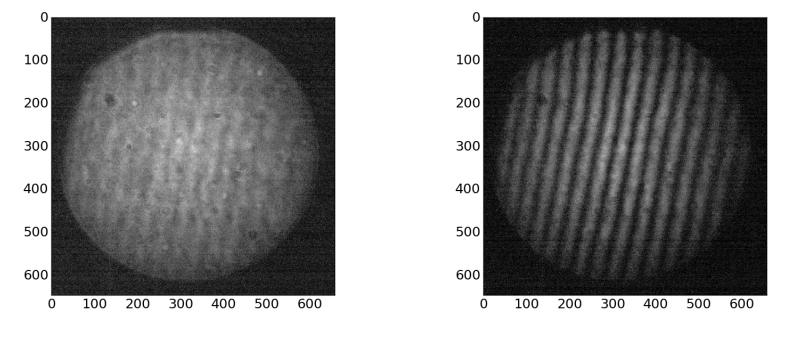}
\caption{Variation in the visibility of fringes in the presence of vibration. {\it Left:} Fringes corresponding to the H$_{\alpha}$ line of the hydrogen lamp taken with one second exposure. The air conditioner was on during the exposure. The vibration caused by the air conditioner  reduced the visibility of the fringes. {\it Right:} Same observation but with air conditioner switched off during the exposure.}
\label{fig:vibrations}
\end{figure}

The Coud\'e room where the TSHS housed was maintained at a temperature of $18^{\circ}$C using an air conditioning unit. However, this has created vibrations in the optical bench where TSHS was kept, since the bench was not floated. Therefore the obtained fringes were of less visibility (Fig.~\ref{fig:vibrations}, {\it Left}), and the visibility degraded with the exposure. Therefore, we had to keep the air-conditioner unit switched off during the TSHS observations.

\subsubsection{Observation of the celestial object}

\begin{figure}[htb!]
\centering
\includegraphics[scale=0.6]{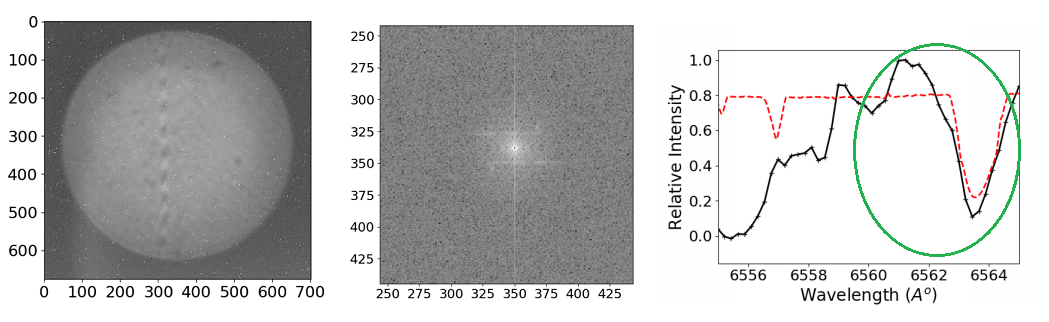}
\caption{{\it Top, Left:} The interferogram obtained for the Betelgeuse. The circular patch is illuminated area of the grating imaged on to the detector. The fringes are visible in the central part of the illuminated area. {\it Middle:} The zoomed version of the 2D Fourier transform of the interferogram  where spectrum is visible. The retrieved spectrum of the star (black solid line) along with the spectrum obtained from ELODIE archive (red dashed line) are plotted together for comparison. The transfer function of the filter is not removed from the spectrum. The wavelength covered (FWHM) by the filter is shown in green circle.}
\label{fig:betel}
\end{figure}

\begin{figure}[h!]
\centering
\includegraphics[scale=0.6]{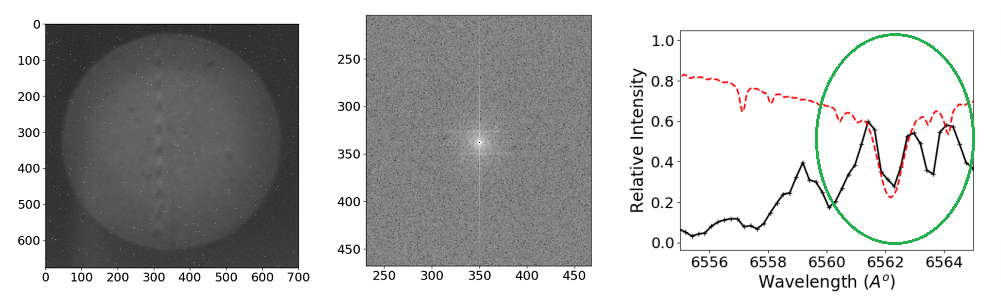}
\caption{{\it Left:} The interferogram obtained for the Sirius. The circular patch is illuminated area of the grating imaged on to the detector. The fringes are visible in the central part of the illuminated area. {\it Middle:} The zoomed version of the 2D Fourier transform of the interferogram  where spectrum is visible. {\it Right:} The retrieved spectrum of the star (black solid line) along with the spectrum obtained from ELODIE archive (red broken line) are plotted together for comparison. The transfer function of the filter is not removed from the spectrum. The wavelength covered (FWHM) by the filter is shown in green circle.}
\label{fig:sirius}
\end{figure}

Due to the low sensitivity of the current detector we could not observe objects fainter than magnitude 1. Therfore, we selected Sirius ($\alpha$ CMa) and Betelgeuse ($\alpha$ Ori) with magnitudes -1.45 and 0.45 (Table.~\ref{tbl:obs}), respectively for the testing of our instrument. 
\begin{table}[htb!]
\caption{Details of Observations conducted on 20 March 2018} 
\label{tbl:obs}
\centering
\begin{tabular}{|l|c|c|c|c|}
\hline
\rule[-1ex]{0pt}{3.5ex}Object & RA & Dec & UT& Exposure(s)\\
\hline
\rule[-1ex]{0pt}{3.5ex}Betelgeuse & 05:55:10& $7^{\circ} 24^{\prime} 26^{\prime \prime}$ &13:41 & 12\\
\hline
\rule[-1ex]{0pt}{3.5ex}Sirius & 06:45:09 & $-16^{\circ} 42^{\prime} 58^{\prime \prime}$ & 13:52& 8\\
\hline
\end{tabular}
\end{table}
A calibration interferogram was taken using the hydrogen lamp before the observations, and then the telescope was pointed towards the source. The resultant interferograms and the two-dimensional Fourier transform of the interferogram for Betelgeuse and Sirius are shown in Left and Middle panels of Figs.~\ref{fig:betel} and \ref{fig:sirius}, respectively. The retrieved spectra, along with the spectra from the ELODIE archive\cite{elodie} for comparison (red dashed line), are shown in the Right panels of Figs.~\ref{fig:betel} and \ref{fig:sirius}, respectively. 

\section{Conclusions}

SHS is a compact, rugged instrument with high resolving power and a narrow but tunable bandpass. In this paper, we have recalled the concept and design of a tunable spatial heterodyne spectrometer in refractive configuration, where we use the beam splitter to split the incoming light. Compared with the all-reflective tunable configuration of SHS the alignment of the instrument is simple since we don't use a roof mirror and fold mirror. The instrument can be tuned to cover a wavelength range from 350 nm to 700 nm with a resolving power of approximately 24000, and is designed for the emission line sources. Based on the Zemax\textsuperscript{\textregistered} simulation of the model, we have procured the components and assembled the instrument on the optical breadboard. We have tested the instrument using a sodium vapor lamp and a halogen lamp monochromator. The spectrum of the sodium vapor lamp was retrieved from the interferogram. We have tested our flat-fielding algorithm on the interferogram with good results. We have also carried out ground-based observations using the TSHS with the optical telescope (VBT), and developed a Python-based pipeline for spectral retrieval. The main objective of the instrument is to study the H$_{\alpha}$ lines from red dwarfs and Be stars. However, with our current detector we could only observe bright stars, such as Sirius and Betelgeuse. In future, we are planning to improve the instrument using better detector. This instrument can be modified to UV regime to fly on a balloon or satellite platform as a field-widened monolithic SHS\cite{watchorn}, which will require additional studies on field-widening techniques.

\acknowledgments 

We would like to thank Prof. Rajpal Singh Sirohi of Tezpur University, for his valuable suggestions and Dr. Binukumar Gopalakrishnan of Indian Institute of Astrophysics and Ajin Prakash of Arksa research lab for their suggestions and contributions in mechanical design of the instrument. Part of this research has been supported by the Department of Science and Technology (Government of India) under the Grant IR/S2/PU-006/2012. A.~G.~Sreejith acknowledges the Austrian  Forschungsförderungsgesellschaft FFG project “ACUTEDIRNDL” P859718.


\bibliography{report}   
\bibliographystyle{spiejour}   

\vspace{1ex}
\noindent Biographies and photographs of the other authors are not available.

\listoffigures
\listoftables

\end{spacing}
\end{document}